\documentclass[aps,pra,twocolumn,10pt,showpacs,superscriptaddress,floatfix]{revtex4-1}
\usepackage[T1]{fontenc}
\usepackage[english]{babel}
\usepackage{amsmath,amssymb,amsbsy,graphicx,braket,mathtools,wrapfig,color,epsfig,import,bm}
\usepackage{todonotes}
\renewcommand{\v}[1]{\ensuremath{\mathbf{#1}}} % for vectors

\newcommand{\D}{^{\dag}}
\newcommand{\p}{^{\prime}}

\newcommand{\conj}{^*}
\newcommand{\MD}{^{\mathstrut}}

\providecommand*{\iu}{\ensuremath{\mathrm{i}}}

\DeclarePairedDelimiter\abs{\lvert}{\rvert}
\newcommand{\vssp}{\vspace{0.1cm}}

\begin{document}
\title{Hall Response and Edge Current Dynamics in Chern Insulators out
  of Equilibrium} \author{M. D. Caio} \affiliation{Department of
  Physics, King's College London, Strand, London WC2R 2LS, United
  Kingdom}
\author{N. R. Cooper} \affiliation{T.C.M. Group, Cavendish Laboratory,
J.J. Thomson Avenue, Cambridge CB3 0HE, United Kingdom}
  \author{M. J. Bhaseen} \affiliation{Department of Physics, King's
    College London, Strand, London WC2R 2LS, United Kingdom}


\begin{abstract}
We investigate the transport properties of Chern insulators following
a quantum quench between topological and non-topological phases.
Recent works have shown that this yields an excited state for which
the Chern number is preserved under unitary evolution. However, this
does not imply the preservation of other physical observables, as we
stressed in our previous work.  Here we provide an analysis of the
Hall response following a quantum quench in an isolated system, with
explicit results for the Haldane model.  We show that the Hall
conductance is no longer related to the Chern number in the
post-quench state, in agreement with previous work.  We also examine
the dynamics of the edge currents in finite-size systems with open
boundary conditions along one direction. We show that the late-time
behavior is captured by a Generalized Gibbs Ensemble, after multiple
traversals of the sample. We discuss the effects of generic open
boundary conditions and confinement potentials.
\end{abstract}

\maketitle

\section{Introduction}

Topological states of matter exhibit a wealth of novel properties due
to their extreme resilience to local perturbations. A striking
manifestation is the quantum Hall effect~\cite{Klitzing1980}, where
the exact quantization of the Hall response is immune to sample
defects~\cite{Laughlin1981a}. This robustness is intimately linked to
the topological Chern invariant, which has direct signatures in
quantum transport~\cite{Thouless1982,Niu1985}. Recent theory and
experiments have exposed many new examples of topological phenomena,
including the relativistic quantum Hall effect in
graphene~\cite{Murthy2003,Novoselov2005a,Zhang2005}, topological
insulators~\cite{Kane2004,Bernevig2006,Konig2007,Khanikaev2013}, and
Majorana edge modes~\cite{Fu2008}. Work has also focused on the
generation of topological Floquet states through time-dependent
driving~\cite{Kitagawa2010,Lindner2011,Rechtsman2013}.

Out of equilibrium, much less is known about the behavior of
topological states. Early works on $p+ip$ superfluids have shown
that topological invariants can be preserved following quenches of the
interaction strength~\cite{Foster2013,Foster2014}. In the context of
Chern insulators, recent studies have shown that the Chern number is
robust under unitary evolution, even following quenches between
topological and non-topological
phases~\cite{DAlessio2015,Caio2015a}. However, as we stressed in
Ref.~\cite{Caio2015a}, this does not imply the persistence of
all physical observables.
%the quantized Hall response.
The presence or absence of edge states for example, depends on the
final Hamiltonian and not just the initial state.  In a similar way,
the Hall response does not necessarily remain quantized.  This is
consistent with recent calculations by Wang {\em et
  al}~\cite{Wang2015,Wang2016}, who have investigated the Hall
response following a quantum quench, including an external reservoir
to induce decoherence and reduce to the diagonal ensemble.

In this manuscript we examine the Hall response following a quantum
quench in a completely isolated system undergoing unitary
evolution. We focus on the Haldane model~\cite{Haldane1988}, as
recently realized using cold atomic gases~\cite{Esslinger2014}. We
show that the Hall response is no longer quantized following a quantum
quench between the topological and non-topological phases, in spite of
the preservation of the Chern index in infinite-size samples. In the
zero frequency limit our results agree with those of
Refs~\cite{Wang2015,Wang2016}, as oscillatory off-diagonal
contributions vanish. The results are in good agreement with
analytical approximations based on the low-energy Dirac
Hamiltonian. We further examine the detailed properties of the edge
currents following a quantum quench in finite-size systems. We show
that the late-time behavior following many traversals of the sample is
quantitatively described by a Generalized Gibbs Ensemble
(GGE)~\cite{Rigol2007,Rigol2008,Caux2012}. With a view towards
  cold atom experiments we also consider the effects of harmonic
  confinement potentials.

\vssp
\begin{figure}
  \centering  
   \includegraphics[width=\columnwidth]{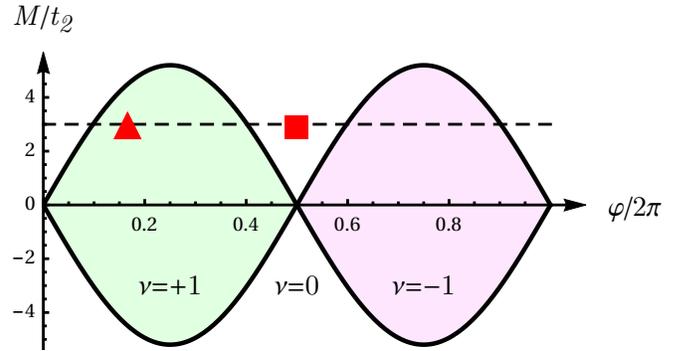}
     \caption{Phase diagram of the Haldane model showing topological
       ($\nu=\pm 1$) and non-topological phases
       ($\nu=0$)~\cite{Haldane1988}. The triangle and the square
       indicate the starting points of quenches
       along the dashed line, as shown in Figs~\ref{fig:cond_omega0a}
       and~\ref{fig:cond_omega0b}.}\label{fig:Hal_Ch}
 \end{figure}

\section{Model}

\begin{figure}
  \centering  
   \includegraphics[width=\columnwidth]{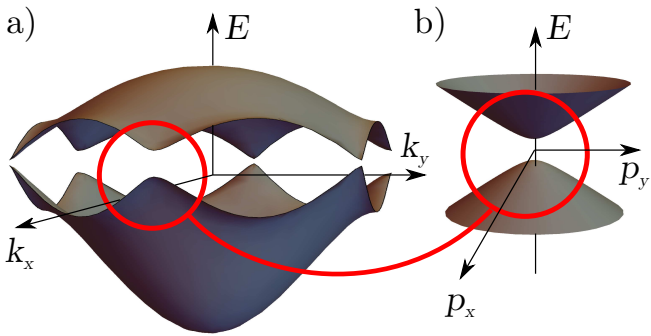}
     \caption{(a) Band structure of the Haldane
       model~\cite{Haldane1988}. (b) At low-energies, the model
       reduces to a sum of two Dirac Hamiltonians with a gapped relativistic
       dispersion relation.}\label{fig:HM_band}
 \end{figure}

The Haldane model describes spinless fermions hopping on a honeycomb
lattice with a staggered magnetic field~\cite{Haldane1988}.  The
Hamiltonian is given by
\begin{eqnarray}
 \hat H&=&-t_1\sum_{\langle i,j\rangle}\left(\hat c\D_i \hat c\MD_j+\mbox{h.c.}\right)
   - t_2 \sum_{\langle\!\langle i,j\rangle\!\rangle}\left(e^{\iu \varphi_{ij}} \hat c\D_i \hat c\MD_j+\mbox{h.c.}\right) \nonumber \\
  & &+M \sum_{i\in A} \hat n_i -M \sum_{i\in B} \hat n_i,\label{eq:HM_realspace}
\end{eqnarray}
where $\hat c_j^\dagger$ and $\hat c_j$ are fermionic creation and
annihilation operators obeying anti-commutation relations $\{\hat
c_j,\hat c_j^\dagger\}=\delta_{ij}$, and $\hat n_i\equiv \hat
c_i^\dagger{\hat c_i}$. Here, $\langle i,j\rangle$ and
$\langle\!\langle i,j\rangle\!\rangle$ indicate summation over the
nearest and next-to-nearest neighbor sites respectively, $t_1$ and $t_2$ are the associated hopping parameters, and $A$ and
$B$ label the two sublattices. The phase $\varphi_{ij}=\pm\varphi$
corresponds to the Aharonov-Bohm phase due to the staggered magnetic
field and is taken positive (negative) in the anti-clockwise
(clockwise) hopping direction. The associated time-reversal symmetry
breaking leads to a quantum Hall effect, in the absence of a net
magnetic field. The energy offset $M$ corresponds to spatial inversion
symmetry breaking, allowing both non-topological and topological
phases to be explored. The phase diagram of the Haldane model is shown
in Fig.~\ref{fig:Hal_Ch}, where we assume that $|t_2/t_1|\le 1/3$ so
that the bands may touch, but not overlap~\cite{Haldane1988}; see Fig.~\ref{fig:HM_band}.
The topological and non-topological phases are distinguished by the
Chern index $\nu$ \cite{Chern1946,Thouless1982,Berry1984,Haldane1988}
which takes the values $\nu=\pm 1$ and $\nu=0$
respectively. This is given by 
\begin{equation}
\nu=\frac{1}{2\pi}\int d^2k \,\Omega,
\label{eq:chernomega}
\end{equation}
where $\Omega=\partial_{k_x}A_{k_y}-\partial_{k_y}A_{k_x}$ is the
Berry curvature, $A_{k_\mu} =i \Braket{\psi|\partial_{k_\mu}|\psi}$ is
the Berry connection and the integral is performed over the first
Brillouin zone. The boundaries of the topological phases correspond to
$M/t_2=\pm 3\sqrt{3} \sin \varphi$, and are independent of $t_1$; see
Fig.~\ref{fig:Hal_Ch}. In the remainder of the manuscript we choose $t_2=t_1/3$
and set $t_1=1$ such that $M$ is expressed in units of $t_1$. We also set 
the intersite spacing $a$ to unity.

\section{Quantum Quenches}

As in Ref.~\cite{Caio2015a}, we study quantum quenches between
different points of the phase diagram shown in
Fig.~\ref{fig:Hal_Ch}. We start from the ground state of $\hat H$ with
parameters $(M_0,\varphi_0)$, and abruptly change them to new values
$(M,\varphi)$.  The system evolves unitarily under the new
Hamiltonian, leading to a time-dependent state
$\Ket{\psi(t)}=\exp\left[-i \hat H(M,\varphi)t/\hbar
  \right]\Ket{\psi_0}$. Here, $\Ket{\psi_0}$ is the initial ground
state,
corresponding to a band insulator of the half-filled system at
$(M_0,\varphi_0)$.  As shown in Refs~\cite{DAlessio2015,Caio2015a},
the corresponding Chern number remains unchanged from its initial
value, even when quenching between different phases. Nonetheless,
other observables may change. In finite-size systems, the topological
and non-topological phases are distinguished by the presence or
absence of edge states. The re-population of these states following a
quantum quench leads to changes in the edge currents and the orbital
magnetization ~\cite{Caio2015a}. This is accompanied by the light-cone
spreading of currents into the interior of the sample, and the onset
of finite-size effects. It is evident that, for a finite-size system,
the concept of a Chern index must eventually break down at late times
following a quantum quench, as one cannot resolve momenta on scales
smaller than the inverse system size, $L^{-1}$ \footnote{We are
  grateful to Prof.~B.~I.~Halperin for this argument.}.
The Berry phase acquired upon circulating a plaquette of size $2\pi/L$ becomes
ill-defined once the Berry connection $A_{k_\mu}$ becomes as large as $|A_{k_\mu}| \sim L$,
and so too does the Chern number. Since, out of equilibrium,
$|\psi\rangle$ is in a superposition of ground and excited states, it
oscillates at the energy difference $\Delta E_{\bm k}$. The
associated Berry connection $A_{k_\mu} = \langle \psi(t) |
\partial_{k_\mu}|\psi(t)\rangle$ grows in time, $t$, as $A_{k_\mu}
\sim [\partial \Delta E_{\bm k}/\partial (\hbar k)_\mu] t = v t$ where $v =
\partial \Delta E_{\bm k}/\partial (\hbar k_\mu)$ is a characteristic
difference of band velocities, limited by $v\lesssim 2c$ with $c$ the
maximum group velocity. Thus $|A_{k_\mu}| \sim L$ when $v t \sim L$,
that is the Chern number becomes ill-defined for timescales larger
than the time required for light-cone propagation across the interior
of the sample, $t\gtrsim L/2c$.

\section{Hall Response}

\begin{figure}
\includegraphics[width=3.2in]{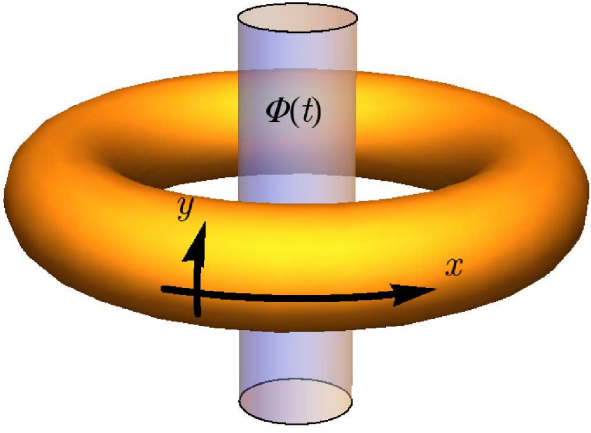}
\caption{Setup used to evaluate the Hall response in the presence of
  periodic boundary conditions. An auxiliary time-dependent magnetic
  flux $\Phi(t)$ generates a longitudinal electric field
  $E_x(t)=-\frac{\partial \Phi(t)}{\partial t}$. Note that this flux
  does {\em not} correspond to the time-reversal symmetry breaking
  parameter $\varphi$ in the Haldane model. The Hall conductance
  $\sigma_{xy}(t)$ is obtained from the transverse current that is in
  phase with $E_x(t)$.}
  \label{fig:torus}
\end{figure}

In order to further differentiate the physical characteristics of the initial
and final states, we examine the Hall response following a quantum
quench. We apply a time-dependent electric field ${\bf E}(t)={\bf
  E}_0\cos\omega t$ and calculate the in-phase transverse response. In the presence of periodic
boundary conditions it is convenient to generate this electric field
by means of an auxiliary magnetic flux threading a toroidal
sample~\cite{Laughlin1981a,Thouless1982,Niu1985}; see 
Fig.~\ref{fig:torus}.
For recent work examining the dynamics of
one-dimensional currents following ``flux quenches'' in ring
geometries see Ref.~\cite{Nakagawa2016}.

In momentum space, the resulting Hamiltonian may be decomposed into a
sum over independent modes $\hat H=\sum_{\bf k} \hat H_{\bf k}$ where
$\hbar \dot{\bf k}=q{\bf E}(t)$, or equivalently $\hbar {\bf k}(t)=\hbar {\bf
  k}(0)+(q{\bf E}_0/\omega)\sin\omega t$, where $q=-e$ is the charge of the carriers. 
 Expanding the Hamiltonian
to linear order in ${\bf E}_0$ yields $\hat H(t)=\sum_{\bf k} \hat
H_{\bf k}+\hat {\bf v}_{\bf k}.(q{\bf E}_0/\omega)\sin\omega t$, where
$\hat {\bf v}_{\bf k}=d\hat H_{\bf k}/\hbar d{\bf k}$ is the velocity
operator. Within the framework of time-dependent perturbation theory
it is convenient to expand the state of the system in the basis of the
unperturbed (${\bf E}_0=0$) post-quench Hamiltonian:
\begin{equation}
\Ket{\psi_{\bf k}(t)}=\sum_{b=l,u} c_{b,{\bf k}}(t) e^{-iE_{b,{\bf
      k}} t/\hbar}\Ket{b,{\bf k}}.
\end{equation}
Here $b=l,u$ refer to the lower
and upper band respectively and $E_{b,{\bf k}}$ are the energies of
the single-particle states; see Fig.~\ref{fig:HM_band}. The case where
the coefficients $c_{b,{\bf k}}$ are constant describes the
unperturbed (${\bf E}=0$) evolution of the system under the
post-quench Hamiltonian. In the presence of ${\bf E}$, first order
perturbation theory yields $c_{b,{\bf k}}(t)\approx c_{b,{\bf
    k}}+\delta c_{b,{\bf k}}(t)$ where
\begin{equation}
  \delta c_{b,{\bf k}}(t)=-\frac{i}{\hbar} \sum_{b^\prime=l,u}\int_0^t dt\p
{\mathcal M}_{b,b^\prime}({\bf k},t^\prime)
  e^{i \Delta_{b,b^\prime}({\bf k})t^\prime}c_{b^\prime,{\bf k}}.\label{eq:delta_c}
\end{equation}
Here, $\Delta_{b,b^\prime}({\bf k})\equiv (E_{b,{\bf
    k}}-E_{b^\prime,{\bf k}})/\hbar$, and ${\mathcal M}_{b,b^\prime}({\bf
  k},t^\prime)\equiv \Bra{b,{\bf k}}\hat {\bf v}_{\bf k}.(q{\bf
  E}_0/\omega)\sin\omega t^\prime\Ket{b^\prime,{\bf k}}$ is the matrix
element of the perturbation. In order to determine the Hall
conductance, we examine the transverse current $\hat J_y=\sum_{\bf
  k}\Braket{\psi_{\bf k}(t)|q\hat v_y|\psi_{\bf k}(t)}$, which flows in
response to an electric field along the $x$-direction. In this pursuit
we neglect the zeroth order contribution which arises in the
absence of $E_0^x$, due the redistribution of carriers between the
bands following the quench.  At first order in $E_0^x$
\begin{equation}
J_y^{(1)}=2 \operatorname{Re}\!\!\left[\sum_{{\bf k}, b,b\p}\!\!\!c_{b,{\bf k}}\conj \delta c_{b\p,{\bf k}}(t) e^{i \Delta_{b,b^\prime}({\bf k})t^\prime} \!\Bra{b,{\bf k}}e\hat v_y \Ket{b\p,{\bf k}}\right]\!. \label{eq:first_order}
\end{equation}
Substituting Eq.~(\ref{eq:delta_c}) into Eq.~(\ref{eq:first_order})
and restricting attention to frequencies below the gap $\omega \ll
\Delta_{b,b^\prime}({\bf k})$, one obtains
\begin{equation}
J_y^{(1)}=-\frac{E_x}{\hbar \omega}\sum_{{\bf k},b,b\p}\operatorname{Re}\left[\mathcal{C}\left(\frac{e^{i\omega t}}{\omega+\Delta_{b,b^\prime}({\bf k})}+\frac{e^{-i\omega t}}{\omega-\Delta_{b,b^\prime}({\bf k})}\right)\right], \label{eq:curr_k}
\end{equation}
where $\mathcal{C}\equiv -i\abs{c_{b,{\bf k}}}^2 \Bra{b,{\bf k}}q \hat
v_y \Ket{b\p,{\bf k}}\Bra{b\p,{\bf k}}q \hat v_x \Ket{b,{\bf k}}$.  In
order to define the Hall conductance we extract the contribution to
$J_y^{(1)}$ that is in phase with the electric field. Denoting
$J_y^{(1)}=I_y\cos\omega t+ \tilde I_y\sin\omega t$ we define
$\sigma_{xy}(\omega)\equiv I_y /AE_x$, where $A$ is the area of the
sample. This yields
\begin{equation}
\sigma_{xy}(\omega)=\frac{q^2}{2\pi h}\sum_{b}
\int d^2k \, \abs{c_{b,{\bf k}}}^2 \tilde\Omega_b(\v k),
\label{eq:sigmaomega}
\end{equation}
where
\begin{equation}
 \tilde\Omega_b(\v k)=-i\sum_{b\p \neq b} \frac{\Bra{b,{\bf k}} \hat v_x \Ket{ b\p,{\bf k} }\Bra{b\p,{\bf k}} \hat v_y \Ket{b,{\bf k}} - \rm{H.c.}}{\omega^2-\Delta_{b,b\p}({\bf k})^2}.
\end{equation}
In the limit $\omega \rightarrow 0$, $\tilde \Omega_b(\v k)$ reduces
to the Berry curvature of the $b$-th band, and the post-quench
d.c. Hall conductance is given by
\begin{equation}
 \sigma_{xy}(0)=\frac{q^2}{2\pi h}\sum_{b} \int d^2k \, \abs{c_{b,{\bf k}}}^2 \Omega_b(\v k), \label{eq:sigmaHall}
\end{equation}
in agreement with Refs~\cite{Wang2015,Wang2016}. An analogous
  result is also found in the context of Floquet systems
  \cite{Dehghani2015}.  Equation (\ref{eq:sigmaHall}) is a
generalization of the Thouless-Kohmoto-Nightingale-den Nijs (TKNN)
formula ~\cite{Thouless1982} to handle (arbitrarily prepared) excited
states. The usual TKNN formula is recovered in the limit where
$|c_{l,{\bf k}}|^2=1$ and $|c_{u,{\bf k}}|^2=0$, corresponding to the
ground state with $\sigma_{xy}(0)=\nu q^2/h$. The result
(\ref{eq:sigmaHall}) has a straightforward physical interpretation in
terms of a semiclassical Boltzmann-like approach. The center of mass
velocity of a wavepacket with momentum ${\bf k}$ and band index $b$ is
given by $\dot{\bf r}_{b,{\bf k}}=\partial\varepsilon_b({\bf k})/\hbar
\partial {\bf k}-(\dot{\bf k}\times \hat z)\Omega_b({\bf k})$, where
$\hbar \dot {\bf k}=q{\bf E}_0$. The second term is the
anomalous velocity associated with the Berry curvature $\Omega_b({\bf
  k})$~\cite{Karplus1954,Kohn1957,Xiao2010a,Chang1995}, where $\hat z$
is a unit vector perpendicular to the sample. The corresponding
current is given by
\begin{equation}
{\bf J}\equiv {\bf J}^{(0)}+{\bf J}^{(1)}=\sum_{b}\int \frac{d^2k}{(2\pi)^2}\,q\dot {\bf r}_{b,{\bf k}}|c_{b,{\bf k}}|^2.
\end{equation}
This yields a contribution proportional to ${\bf E}_0$:
\begin{equation}
{\bf J}^{(1)}=-\frac{q^2}{2\pi h}\sum_{b}\int d^2k \, |c_{b,{\bf k}}|^2({\bf E}_0\times \hat z)\Omega_b({\bf k}),
\end{equation}
in agreement with the Hall response (\ref{eq:sigmaHall}). In
equilibrium, this can be used to extract the
Berry curvature \cite{Price2012} from measurements of the transverse drift~\cite{Esslinger2014,Aidelsburger2014}.

\begin{figure}[h]
  \centering  
   \includegraphics[width=\columnwidth]{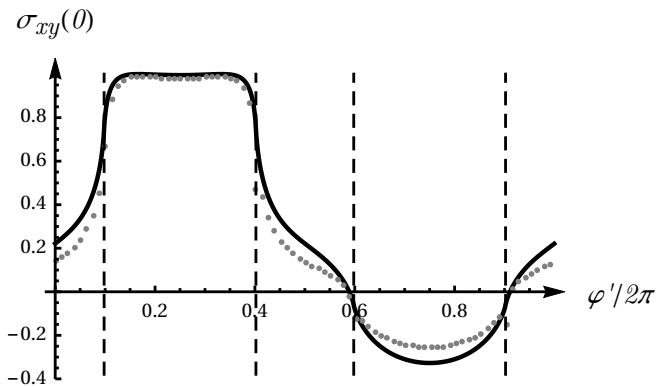}
     \caption{Hall conductance in units of $q^2/h$ following a quantum
       quench from the topological phase with $\varphi=\pi/3$ and
       $M=1$ (as indicated by the triangular symbol in
         Fig.~\ref{fig:Hal_Ch}) to $\varphi=\varphi\p$ and $M=1$. The
       gray dots are numerical results for the Haldane model obtained
       from Eq.~(\ref{eq:sigmaHall}). The black solid line is the
       analytical result for the corresponding quenches in the
       low-energy Dirac approximation, obtained by summing
       contributions from Eq.~(\ref{eq:sigmaDirac}). The results are
       in quantitative agreement for small quenches in the vicinity of
       $\varphi=\pi/3$, and in qualitative agreement for larger
       quenches.  The vertical dashed lines correspond to the
       boundaries of the topological phases. The Hall conductance
       remains numerically close to $q^2/h$ for quenches within the
       same topological phase ($\nu=1$), but does not saturate at
       $-q^2/h$ when quenching to the other topological phase
       ($\nu=-1$).}
       \label{fig:cond_omega0a}
 \end{figure}

\begin{figure}[h]
  \centering  
   \includegraphics[width=\columnwidth]{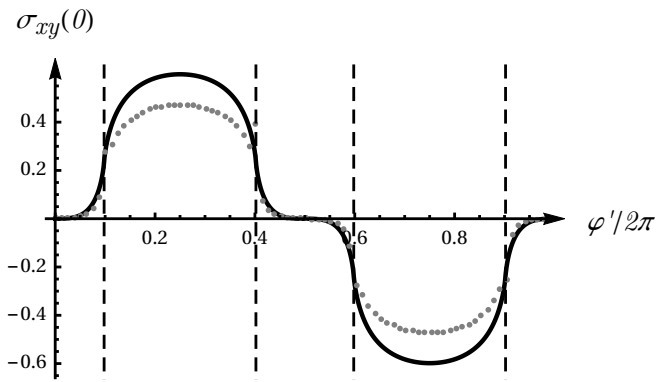}
     \caption{Hall conductance in units of $q^2/h$ following a quantum
       quench from the non-topological phase with $\varphi=\pi$ and
       $M=1$ (as indicated by the square symbol in
         Fig.~\ref{fig:Hal_Ch}) to $\varphi=\varphi\p$ and $M=1$. The
       gray dots are numerical results for the Haldane model obtained
       from Eq.~(\ref{eq:sigmaHall}). The black solid line is the
       analytical result for the corresponding quenches in the
       low-energy Dirac approximation, obtained by summing
       contributions from Eq.~(\ref{eq:sigmaDirac}). The results are
       in quantitative agreement for small quenches in the vicinity of
       $\varphi=\pi$, and in qualitative agreement for larger
       quenches.  The vertical dashed lines correspond to the
       boundaries of the topological phases.}\label{fig:cond_omega0b}
 \end{figure}

\begin{figure}
\includegraphics[width=3.2in]{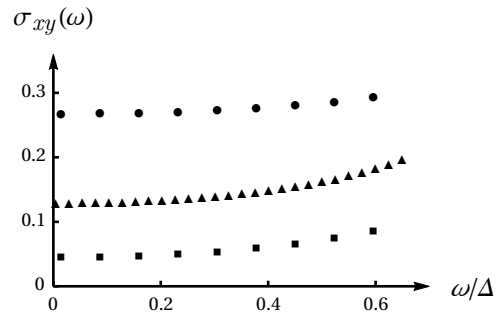}
\caption{Hall response $\sigma_{xy}(\omega)$ in units of $q^2/h$ evaluated using
  Eq.~(\ref{eq:sigmaomega}) for frequencies below the  direct band gap $\Delta \equiv{\rm min}[\Delta_{b,b\p}({\bf k})]$, following a quantum quench from the
  topological phase with $M=1$ and $\varphi=\pi/3$, to the
  non-topological phase with $M=1$ and $\varphi=\varphi^\prime$. 
The three curves correspond to $\varphi\p=7\pi/8$ (circles),
  $\varphi\p=\pi$ (triangles) and $\varphi\p=9\pi/8$ (squares), for which
 $\Delta \sim
  0.67$, $2$, and $0.67 \times t_1/ \hbar$, respectively.}
  \label{fig:cond_omega}
\end{figure}

In Fig.~\ref{fig:cond_omega0a} we show numerical results for the Hall
conductance in the post-quench state, starting from the topological
phase. It is readily seen that the values are no longer quantized in
integer multiples of $q^2/h$. Similarly, quenches from the
non-topological to the topological phase also fail to yield the
quantized values found in the equilibrium ground state with $\nu=\pm
1$; see Fig.~\ref{fig:cond_omega0b}. Heuristically, the quench ``heats
up'' the sample by promoting carriers to the upper band which
contribute to the Hall response by the Berry curvature of that band,
thus yielding a non-quantized Hall response.
More generally, we may use
Eq.~(\ref{eq:sigmaomega}) to determine the a.c. Hall response for
frequencies smaller than the band gap, $\omega\ll \Delta_{b,b^\prime}({\rm
  k})$; see Fig.~\ref{fig:cond_omega}.
For larger frequencies, non-diagonal contributions of the form
$c_{b,{\bf k}}c_{b^\prime,{\bf k}}^\ast$ do not disappear under
unitary evolution and the current (\ref{eq:first_order}) must be
employed.

\subsection{Low-Energy Approximation}

\begin{figure}[h]
  \centering \includegraphics[width=\columnwidth]{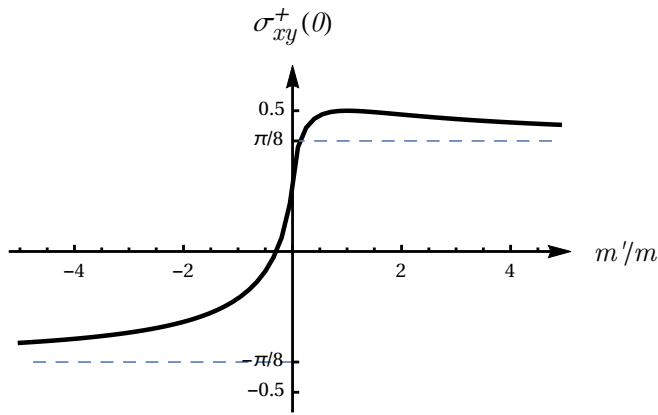}
     \caption{Post-quench Hall conductance in units of $q^2/h$ for a
       single Dirac point ($\alpha=+$), following a quench from $m>0$
       to $m\p$.  For the same choice of quench parameters, the
       other Dirac point ($\alpha=-$) displays the opposite Hall
       conductance. Note that for quenches in the Haldane model, the
       changes in mass for each Dirac point may differ, yielding a
       non-zero Hall response. The dashed lines correspond to
       the asymptotes $\sigma_{xy}(0)=\pm(\pi/8)q^2/h$, which can be determined
       analytically from
       Eq.~(\ref{eq:sigmaDirac}).}\label{fig:cond_Dirac}
 \end{figure}

Analytical results for the Hall response (\ref{eq:sigmaHall}) may be
obtained within the framework of the low-energy Dirac Hamiltonian for
sufficiently small quenches. At low-energies, the Haldane model may be
described as the sum of two Dirac Hamiltonians $H=H_++H_-$
~\cite{Haldane1988} where
\begin{equation}
H_\alpha = \begin{pmatrix}
            m_\alpha c^2 && -c p e^{i\alpha \theta} \\ -c p e^{-i \alpha \theta} && -m_\alpha c^2
           \end{pmatrix}.
\end{equation}
Here, $\alpha=\pm$ labels the two inequivalent Dirac points,
$c=3t_1 a/2\hbar$ is the effective speed of light, $p \exp(i \theta)$
parametrizes the 2D momentum ($p_x$,$p_y$) and
$m_\alpha=(M-3\sqrt{3}\alpha t_2 \sin \varphi)/c^2$ is the effective
Dirac fermion mass~\cite{Haldane1988}. In this representation, a
quench of the Haldane parameters $M\rightarrow M\p$ and
$\varphi\rightarrow \varphi\p$ corresponds to quenches of the
effective masses $m_\alpha \rightarrow m_\alpha\p$. Combining results
for the Berry curvature~\cite{Price2012}, with the Dirac band
occupation following a quench~\cite{Caio2015a}, one obtains
\begin{equation}
 \sigma_{xy}^\alpha(0)=\frac{\alpha e^2}{2 h}\int_0^\infty dp\frac{cpm_\alpha\p(p^2+c^2 m_\alpha m_\alpha\p)}{\sqrt{p^2+c^2 m_\alpha^2}{(p^2+c^2 {m_\alpha\p}^2)}^2}, \label{eq:sigmaDirac}
\end{equation}
for the two Dirac points. This is in agreement with
Refs~\cite{Wang2015,Wang2016}. In deriving this result, the integral
over the two-dimensional Brillouin zone has been replaced by an
integral over the infinite two-dimensional momentum-space. The solid
lines in Figs~\ref{fig:cond_omega0a} and ~\ref{fig:cond_omega0b}
correspond to the low-energy approximation
$\sigma_{xy}(0)=\sigma_{xy}^+(0)+\sigma_{xy}^-(0)$. The approximation
is in good agreement with the exact numerical results for the lattice
model (\ref{eq:HM_realspace}). For small quenches, which do not explore the non-linear
regime of the dispersion relation, the results agree
quantitatively. For larger quenches, the approximation breaks down,
but it still captures the qualitative features. For large quenches,
with $\abs{m-m\p}\rightarrow \infty$, the contribution to the Hall
conductance from a single Dirac point shows plateaus at
$(\pm\pi/8)q^2/h$; see Fig.~\ref{fig:cond_Dirac}.  These results are
consistent with the notion that preservation of $\nu$ does {\em not}
imply the preservation of the Hall response $\sigma_{xy}(0)$. In the
absence of interactions, the final state is non-thermal, and characterized by the occupations $c_{b,{\bf
    k}}$ in Eq.~(\ref{eq:sigmaHall}). In this excited state the Hall response need not be quantized.

\section{Generalized Gibbs Ensemble}

\begin{figure}[h]
  \centering  
   \includegraphics[width=\columnwidth]{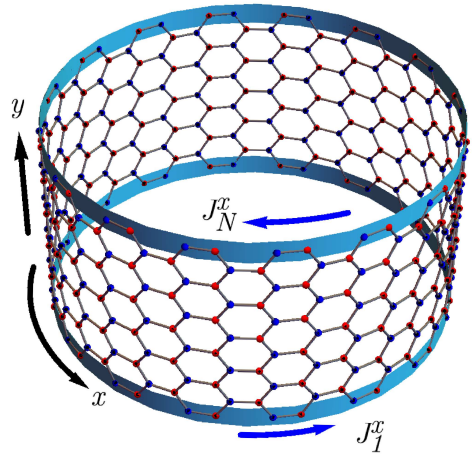}
     \caption{Finite-size cylindrical geometry with periodic (open)
       boundary conditions along the $x$- ($y$-) direction. When
       $\varphi\neq 0$ there are generically edge currents, $J_1^x$
       and $J_N^x$, flowing along the sample boundaries in opposite
       directions. After a quantum quench, the edge currents evolve as
       a function of time, and currents flow into the interior of the
       sample.}\label{fig:cyl_edg}
 \end{figure}

A quantitative description of the non-thermal post-quench state can be
obtained by considering the system in a cylindrical geometry with open
boundary conditions along one direction; see
Fig.~\ref{fig:cyl_edg}. In our previous work we showed that the
resulting edge currents undergo non-trivial dynamics, due to the
presence or absence of edge modes in the spectrum \cite{Caio2015a}.
\begin{figure}
  \centering  
   \includegraphics[width=\columnwidth]{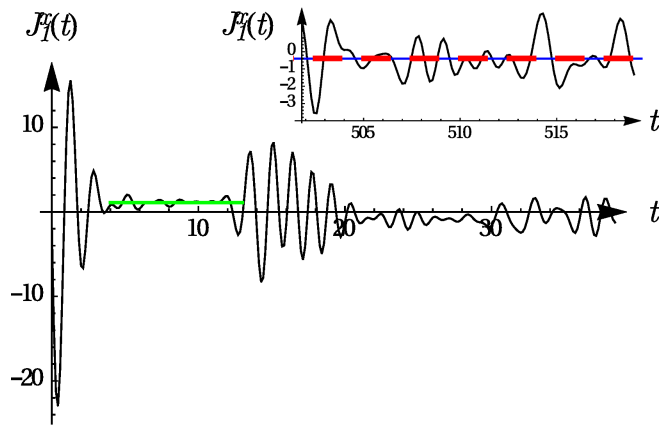}
     \caption{(Main) Total edge current $J_1^x(t)$ in units of $q t_1 a / \hbar$
       following a quantum
       quench within the topological phase with $\nu=-1$. The time $t$
       is measured in units of $\hbar/t_1$. We set
       $\varphi=\pi/3$ and quench $M$ from $1.4$ to $-1.4$. The solid
       horizontal (green) line corresponds to the time-averaged total
       edge current, in the quasi-stationary regime before the onset
       of finite-size traversals. (Inset) Late-time data after $29$
       traversals of the sample. The dashed (red) line corresponds to
       the prediction of the GGE. This agrees with the time-average of
       the late-time data, as shown by the solid horizontal line
       (blue), to within approximately $3\%$.}\label{fig:gge}
 \end{figure}
\begin{figure}
  \centering  
   \includegraphics[width=\columnwidth]{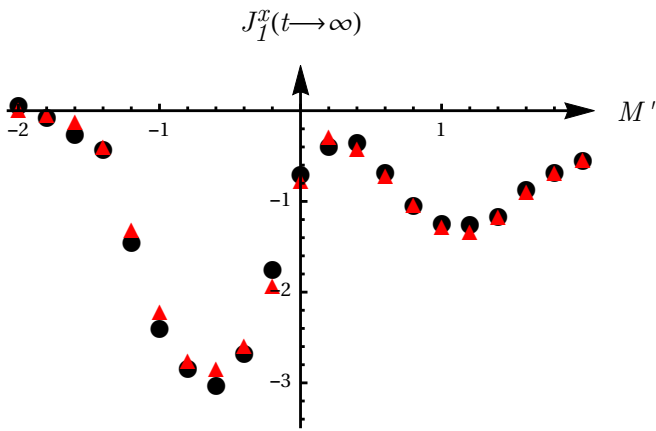}
   \caption{Comparison of the time-averaged total edge current in the
     Haldane model at late times (circles) and the prediction
     $\Braket{\hat\rho_{{\rm GGE}}\hat J^x_1}$ of the GGE (triangles)
     for quantum quenches with $\varphi=\pi/3$ held fixed and
     $M=1.4\rightarrow M\p$. 
     The time averaging is performed over a
     single traversal period of the finite-size system, following $29$
     traversals of the sample.
     The currents are measured in units of $q t_1 a / \hbar$.}\label{fig:GGEvsTimeAv}
 \end{figure}
In particular, the edge currents decay towards new
values that depend on the post-quench Hamiltonian, and not just the
initial state; see Fig.~\ref{fig:gge}. This evolution is accompanied
by light-cone spreading of currents into the interior of the sample,
with the eventual onset of finite-size effects and resurgent
oscillations. Here, we show that the long time behavior of the
total edge current, after many traversals of the sample, is
captured by a Generalized Gibbs Ensemble (GGE)
\cite{Rigol2007,Rigol2008,Caux2012}. The density matrix is given by
\begin{equation}
\hat\rho_{{\rm GGE}} = Z^{-1} \exp{\left(-\sum_{\gamma} \lambda_{\gamma}
  \hat{n}_\gamma\right)},
\label{eq:GGE}
\end{equation}
where $\hat n_\gamma$ are the conserved occupations of a given energy
state, $\gamma$ labels the energy level and the momentum index, and $Z={\rm
  Tr}\exp{\left(-\sum_{\gamma} \lambda_{\gamma} \hat{n}_\gamma\right)}$; see Appendix
\ref{app:C}. Here, the $\lambda_\gamma$'s are generalized inverse
temperatures. These are determined by the self-consistency condition
that the mode occupations immediately after the quench coincide with
the averages computed via the GGE, $\Braket{\hat n_\gamma}={\rm
  Tr}(\hat\rho_{{\rm GGE}} \hat n_\gamma)$. This yields
$\lambda_\gamma=\ln[(1-\Braket{\hat n_\gamma})/\Braket{\hat n_\gamma}]$~\cite{Rigol2007}. 
In Fig.~\ref{fig:GGEvsTimeAv} we show the comparison
between the time-averaged total edge current along a single edge at
late times, and the predictions of the GGE. The results are in
excellent agreement.

\section{Confinement Potentials}

Thus far, we have explored the dynamics of the Haldane model in
toroidal and cylindrical geometries, as periodic boundary conditions
provide considerable simplifications for theory and simulation. In
order to make clear predictions for experiment, it is instructive to
consider finite-size samples with open boundaries, especially due to
the use of optical traps for cold atoms.  We first consider the
effects of transverse confinement in the cylindrical geometry depicted
in Fig.~\ref{fig:cyl_edg}, before discussing the case of rotationally
symmetric confinement. For earlier work exploring the effects of
  trapping potentials on the equilibrium edge physics of topological 
  systems see Refs~\cite{Stanescu2010,Buchhold2012}.

\begin{figure}[ht]
 \includegraphics[width=\columnwidth]{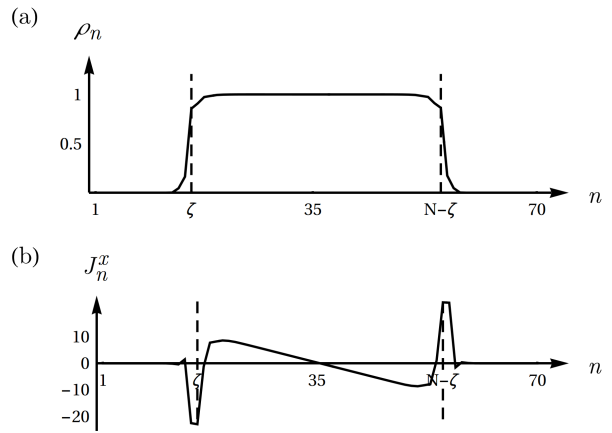}
 \caption{Equilibrium properties of the Haldane model in the
   cylindrical geometry shown in Fig.~\ref{fig:cyl_edg}, with an
   additional harmonic confinement potential $V_n$ given by
   Eq.~\ref{eq:vn}. We consider the topological phase with $M=0$ and
   $\varphi=\pi/3$ and set $N=70$, $\mathcal{C}=1.2$ and
   $\zeta=15$. (a) Particle density $\rho_n$ as a function of the row
   index $n\in 1,...,70$. At half-filling, the potential confines the
   particles into the region $\zeta<n<N-\zeta$, as indicated by the
   vertical dashed lines. This imposes an effective system size in
   which it is possible to observe edge effects. (b) Total
   longitudinal current $J_n^x$ along the cylinder showing clearly
   separated edge currents, broadened due to the smooth 
   potential. Hall currents exist in the interior of the sample due to
   the effective electric field generated by the trap.}
 \label{fig:DandJ}
\end{figure}

We consider the Haldane model in the setup shown in
Fig.~\ref{fig:cyl_edg}, with an additional harmonic confinement
potential applied along the $y$-direction, $V(y)\propto (y-y_0)^2$.
Here $y_0$ is the center of the trap which we take to be located at
the midpoint of the strip. It is convenient to parameterize
the potential as
\begin{equation}
  V_n={\mathcal C}\left(\frac{n-N/2}{\zeta-N/2}\right)^2,
  \label{eq:vn}
\end{equation}
where $n=1,\dots,N$ labels the row of the strip, $\zeta$ controls the
effective width of the confined sample, and ${\mathcal C}$ is a
constant. For suitable choices of ${\mathcal C}$ and $\zeta$, for a
strip of a given width $N$, the equilibrium particle density is
approximately uniform for $\zeta< n< N-\zeta$; see
Fig.~\ref{fig:DandJ}(a). The corresponding current profile in the
topological phase shows clearly separated edge currents that are
broadened due to the smooth confinement potential; see
Fig.~\ref{fig:DandJ}(b). In contrast to the case of hard wall
boundaries \cite{Caio2015a}, the bulk also contains longitudinal Hall
currents if it corresponds to a topological phase with non-zero Chern
number. These arise due to the effective transverse electric field
generated by the harmonic potential.

\begin{figure}[ht]
 \includegraphics[width=\columnwidth]{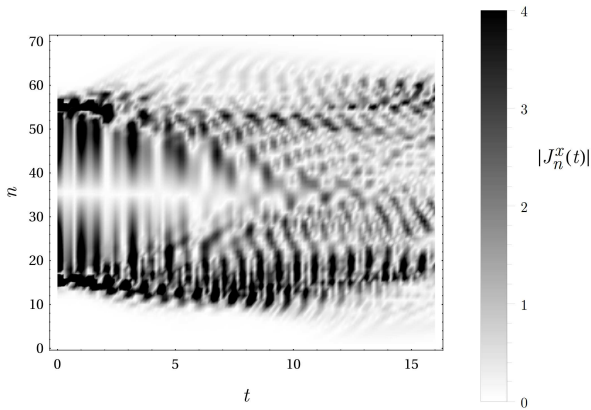}
 \caption{Dynamics of the currents $|J_n^x(t)|$ following a quantum
   quench from the topological phase with $M=0$ and $\varphi=\pi/3$ to
   the non-topological phase $M=3$ and $\varphi=\pi/3$. We set $N=70$,
   $\mathcal{C}=1.2$ and $\zeta=15$. The spreading of the currents
   into the interior (and to a lesser extent the exterior) of
   the effective sample is visible even in the presence of the
   harmonic potential. However, the propagation speed departs from the
   effective speed of light, due to the broadening of the edge
   currents and the contribution of bulk Hall currents.}
 \label{fig:spreading}
\end{figure}

Following a quantum quench, the edge currents spread towards the
interior of the sample, as found in the case of hard wall boundaries
\cite{Caio2015a}; see Fig.~\ref{fig:spreading}. The currents
  also spread outside the initial sample area to a lesser extent due
  to the harmonic confinement. The light-cone propagation is
clearly visible in the presence of the trap, but the apparent speed of
light differs slightly from the uniform case. This is attributed to
the broadening of the edge current profiles and the presence of the
equilibrium bulk Hall currents, induced by the harmonic potential.

\begin{figure}[ht]
 \includegraphics[width=2.9 in]{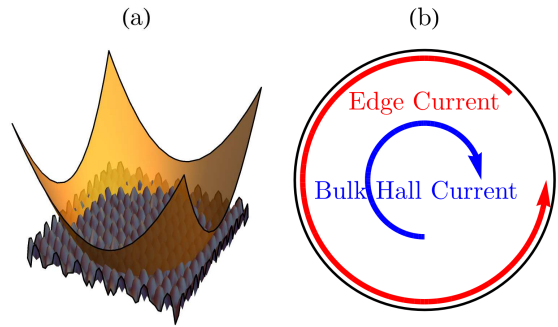}
 \caption{(a) Illustration of the potential landscape of a hexagonal
   optical lattice in the presence of a rotationally symmetric
   harmonic trap. (b) Circulating equilibrium currents in the
   resulting disk geometry. Following a quantum quench between the
   topological and the non-topological phases the edge currents are
   expected to propagate into the interior (and the exterior) of
   the sample at the effective speed of light. This will be smoothed
   out due to the broadening of the edge currents and the presence of
   the bulk Hall currents. }
 \label{fig:circular}
\end{figure}

The case of fully open boundary conditions presents a significant
increase in the numerical computations required, but is not expected
to lead to different results. In the presence of a rotationally
symmetric harmonic trap, as shown in Fig.~\ref{fig:circular}(a),
broadened edge currents will flow on the effective sample boundaries;
see Fig.~\ref{fig:circular}(b). Likewise, equilibrium bulk Hall
currents will also circulate (in the opposite sense to the edge
currents) due to the effective radial electric field. Following a
quantum quench, the edge currents will flow towards the interior of
the sample, exhibiting light-cone propagation; this will be smoothed
out by the edge current broadening and the bulk Hall currents, which
are present even in equilibrium.

An estimate of the timescales for the light-cone propagation can be
inferred from the experimental parameters used in
Ref.~\cite{Esslinger2014}. Using the typical hopping parameter
  $t_1\sim 10^2\,h${\rm Hz}, the effective speed of light is $c\sim
  10^3\,a/\rm{s}$, measured in intersite spacings per
  second. i.e. lattice site propagation occurs on millisecond
  timescales. The timescale for observing the onset of the 
  current plateau in Fig.~\ref{fig:gge}, and the light-cone
  propagation in Fig.~\ref{fig:spreading}, is thus of the order of
  several milliseconds; the unit of time in Figs~\ref{fig:gge} and
  \ref{fig:spreading} is $\hbar/t_1\sim 10^{-3}\,\rm{s}$. This is
  comparable with the timescales explored in experiment
  \cite{Esslinger2014}.
  
\section{Conclusions}

In this paper we have explored the transport properties of Chern
insulators following a quantum quench in an isolated system undergoing
unitary evolution. The Hall conductance is no longer described by the
ground state relation $\sigma_{xy}(0)=\nu q^2/h$, in spite of the
preservation of $\nu$ in infinite-size systems; the Chern index is a
property of the state, whilst the Hall response depends both on the
state and the final Hamiltonian.  In the presence of open boundary
conditions, the total edge currents are described by a GGE at late
times. It would be interesting to explore the quench dynamics of Chern
insulators in experiment, probing their Hall response and edge current
dynamics.

\section*{Acknowledgments}

We acknowledge helpful conversations with F. Essler, N. Goldman and
B. Halperin.  This work was supported by EPSRC Grants EP/J017639/1 and
EP/K030094/1. MJB thanks the EPSRC Centre for Cross-Disciplinary
Approaches to Non-Equilibrium Systems (CANES) funded under grant
EP/L015854/1. MJB and MDC thank the Thomas Young Center.

\vspace*{0.2cm}

\vssp \emph{Note added.---} Whilst this work was being finalized for
publication the pre-prints \cite{Hu2016,Wilson2016} appeared which
also investigate the Hall response out of equilibrium.

\vspace*{0.2cm}

%%%%%%%%%%%%%%%%%%%%%%%%%%%%%%%%%%%%%%%%%%%%%%%%%%%%%%%%%%
%%%%%%%%%%%%%%%%%%%%%%%%%%%%%%%%%%%%%%%%%%%%%%%%%%%%%%%%%%
%%%%%%%%%%%%%%%%%%%%%%%%%%%%%%%%%%%%%%%%%%%%%%%%%%%%%%%%%%
%%%%%%%%%%%%%%%%%%%%%%%%%%%%%%%%%%%%%%%%%%%%%%%%%%%%%%%%%%
%%%%%%%%%%%%%%%%%%%%%%%%%%%%%%%%%%%%%%%%%%%%%%%%%%%%%%%%%%
%%%%%%%%%%%%%%%%%%%%%%%%%%%%%%%%%%%%%%%%%%%%%%%%%%%%%%%%%%
%%%%%%%%%%%%%%%%%%%%%%%%%%%%%%%%%%%%%%%%%%%%%%%%%%%%%%%%%%
%%%%%%%%%%%%%%%%%%%%%%%%%%%%%%%%%%%%%%%%%%%%%%%%%%%%%%%%%%
%%%%%%%%%%%%%%%%%%%%%%%%%%%%%%%%%%%%%%%%%%%%%%%%%%%%%%%%%%

%

%%%%%%%%%%%%%%%%%%%%%%%%%%%%%%%%%%%%%%%%%%%%%%%%%%%%%%%%%%
%%%%%%%%%%%%%%%%%%%%%%%%%%%%%%%%%%%%%%%%%%%%%%%%%%%%%%%%%%
%%%%%%%%%%%%%%%%%%%%%%%%%%%%%%%%%%%%%%%%%%%%%%%%%%%%%%%%%%
%%%%%%%%%%%%%%%%%%%%%%%%%%%%%%%%%%%%%%%%%%%%%%%%%%%%%%%%%%
%%%%%%%%%%%%%%%%%%%%%%%%%%%%%%%%%%%%%%%%%%%%%%%%%%%%%%%%%%
%%%%%%%%%%%%%%%%%%%%%%%%%%%%%%%%%%%%%%%%%%%%%%%%%%%%%%%%%%
%%%%%%%%%%%%%%%%%%%%%%%%%%%%%%%%%%%%%%%%%%%%%%%%%%%%%%%%%%
%%%%%%%%%%%%%%%%%%%%%%%%%%%%%%%%%%%%%%%%%%%%%%%%%%%%%%%%%%
%%%%%%%%%%%%%%%%%%%%%%%%%%%%%%%%%%%%%%%%%%%%%%%%%%%%%%%%%%
%%%%%%%%%%%%%%%%%%%%%%%%%%%%%%%%%%%%%%%%%%%%%%%%%%%%%%%%%%

\onecolumngrid

\vspace{0.5cm}

%\pagebreak

\appendix
\twocolumngrid

\section{Low-Energy Approximation} \label{app:A}

In Figs~\ref{fig:cond_omega0a} and \ref{fig:cond_omega0b} we compare the
post-quench Hall response in the Haldane model (\ref{eq:HM_realspace})
with the low-energy Dirac approximation (\ref{eq:sigmaDirac}). For
completeness we provide a derivation of the latter; see also
Refs~\cite{Wang2015,Wang2016}.  For a single Dirac point, the
occupation probability amplitudes of the lower and upper bands
following an effective mass quench $m\rightarrow m\p$, are given
by~\cite{Caio2015a}
\begin{equation}\begin{split}
 a(m,m\p,p)&=f_-(p,m)f_-(p,m\p)+f_+(p,m)f_+(p,m\p),\\
 b(m,m\p,p)&=f_+(p,m)f_-(p,m\p)-f_-(p,m)f_+(p,m\p),
\end{split}\end{equation}
respectively, where $f_\pm(p,m)\equiv
\sqrt{\frac{1}{2}\left(1\pm\frac{mc}{\sqrt{p^2+m^2
      c^2}}\right)}$. Here we have parameterized the two-dimensional
momentum space with $(p,\theta)$, as in the main text.  For a single
Dirac point $\alpha$, the Berry curvature of the upper band is given
by $\alpha \Omega_{\rm D}(p,m)$, where $\Omega_{\rm
  D}(p,m)=cpm/(p^2+m^2 c^2)^{3/2}$; the Berry curvature of the lower
band has the opposite sign~\cite{Price2012}. Substituting these
results into Eq.~(\ref{eq:sigmaHall}) it follows that after a quench
$m_\alpha\rightarrow m_\alpha\p$,
\begin{multline}
 \sigma_{xy}^\alpha(0)=\frac{\alpha q^2}{2 h}\int_0^\infty dp\,\Omega_{\rm D}(p,m_\alpha\p)\, \times \\ \left(\abs{b(m_\alpha,m_\alpha\p,p)}^2-\abs{a(m_\alpha,m_\alpha\p,p)}^2\right).
\end{multline}
Eq.~(\ref{eq:sigmaDirac}) follows straightforwardly~\cite{Wang2015,Wang2016}.

\section{Edge Currents}\label{app:B}

In order to study the dynamics of the edge currents, we consider the
Haldane model (\ref{eq:HM_realspace}) in a finite-size cylindrical
geometry, with periodic (open) boundary conditions along the $x$- ($y$-)
direction; see Fig.~\ref{fig:cyl_edg}. We define the local current
flowing through a site $l$ of the lattice as~\cite{Caio2015a}
\begin{equation}
 \hat{\mathbf{J}}_l :=-\frac{iq}{2}\sum_j \boldsymbol \delta_{jl}(t_{lj} \hat c_l\D \hat c_j - {\rm H}.{\rm c}.),
\end{equation}
where $t_{lj}$ is the hopping parameter, $\boldsymbol \delta_{jl}$ is
the vector joining site $j$ to $l$, and the sum is performed over
nearest and next-nearest neighbors. Each lattice index corresponds to
a triplet of indices $\{m=1,...,M;n=1,...,N;s=A,B\}$, labeling the
$x$- and $y$- positions of the unit cell, and the
sublattice~\cite{Caio2015a}. The total longitudinal current flowing
along the lower edge of the cylinder in the $x$-direction is given by
$J_1^x=\langle \hat J_1^x\rangle =\sum_{ms}\langle \hat
J^x_{m1s}\rangle$, as shown in Figs~\ref{fig:cyl_edg} and
\ref{fig:gge}.

\section{Generalized Gibbs Ensemble}\label{app:C}
In the finite-size cylindrical geometry depicted in
Fig.~\ref{fig:cyl_edg}, the Hamiltonian can be diagonalized as $\hat
H= \sum_{\Upsilon,k_x}\epsilon_\Upsilon(k_x){\hat f}\D_\Upsilon(k_x){\hat
  f}\MD_\Upsilon(k_x)$, where we have exploited the periodicity in the
$x$-direction, and $\Upsilon=1,\dots,2N$ labels the energy levels at each
$k_x$-point. In the non-interacting model
(\ref{eq:HM_realspace}) the number operators $\hat n_\Upsilon(k_x)={\hat
  f}\D_\Upsilon(k_x){\hat f}\MD_\Upsilon(k_x)$ are conserved. Denoting the
pair of indices $\{\Upsilon,k_x\}$ by $\gamma$, the late-time values of the
time-averaged edge currents are described by the GGE given in
Eq.~(\ref{eq:GGE}).

\onecolumngrid

\end{document}